\documentclass[letterpaper,twocolumn,10pt]{article}
\usepackage{usenix2024_SOUPS}

\usepackage{mdframed}
\usepackage{booktabs}
\usepackage{array}
\usepackage{graphicx}
\usepackage{appendix}
\usepackage{tikz}
\usepackage{amsmath}

\newcommand{\newadded}[1]{\textcolor{black}{#1}}
\begin{document}
\date{}
%\title{``Hello, is this Anna?": A First Look at Pig-Butchering Scams}
\title{``Hello, is this Anna?": Unpacking the Lifecycle of Pig-Butchering Scams}

%for single author (just remove % characters)
%{\rm Anonymous Authors}}
% % copy the following lines to add more authors
% % \and
% % {\rm Name}\\
% %Name Institution
% } % end author
\author{
Rajvardhan Oak \\
University of California Davis\\
rvoak@ucdavis.edu
\and 
Zubair Shafiq \\
University of California Davis\\
zubair@ucdavis.edu
}

\maketitle
\thecopyright

\begin{abstract}
Pig-butchering scams have emerged as a complex form of fraud that combines elements of romance, investment fraud, and advanced social engineering tactics to systematically exploit victims.
In this paper, we present the first qualitative analysis of pig-butchering scams, informed by in-depth semi-structured interviews with $N=26$ victims. We capture nuanced, first-hand accounts from victims, providing insight into the lifecycle of pig-butchering scams and the complex emotional and financial manipulation involved. 
We systematically analyze each phase of the scam, revealing that perpetrators employ  tactics such as staged trust-building, fraudulent financial platforms, fabricated investment returns, and repeated high-pressure tactics, all designed to exploit victims’ trust and financial resources over extended periods.
Our findings reveal an  organized scam lifecycle characterized by emotional manipulation, staged financial exploitation, and persistent re-engagement efforts that amplify victim losses. 
We also find complex psychological and financial impacts on victims, including heightened vulnerability to secondary scams. Finally, we propose actionable intervention points for social media and financial platforms to curb the prevalence of these scams and highlight the need for non-stigmatizing terminology to encourage victims to report and seek assistance.
\end{abstract}

% \begin{abstract}
% Pig-butchering scams, or Sha Zhu Pan, have emerged as a complex form of cyber-enabled financial fraud that combines elements of romance, investment fraud, and advanced social engineering tactics to systematically exploit victims.
% %
% In this paper, we present the first qualitative analysis of pig-butchering scams, informed by in-depth semi-structured interviews with $N=26$ victims. We capture nuanced, first-hand accounts from victims across multiple regions, providing insight into the lifecycle of pig-butchering scams and the complex emotional and financial manipulation involved. 
% %
% We systematically analyze each phase of the scam, revealing that perpetrators employ  tactics such as staged trust-building, fraudulent financial platforms, fabricated investment returns, and repeated high-pressure tactics, all designed to exploit victims’ trust and financial resources over extended periods.
% %
% Our findings reveal an  organized scam lifecycle characterized by emotional manipulation, staged financial exploitation, and persistent re-engagement efforts that amplify victim losses. 
% We also find complex psychological and financial impacts on victims, including heightened vulnerability to secondary scams. Finally, we propose actionable intervention points for social media and financial platforms to curb the prevalence of these scams and highlight the need for non-stigmatizing terminology to encourage victims to report and seek assistance.
% \end{abstract}


\section{Introduction}

\begin{figure}[tb]
\centering
\includegraphics[width=.25\textwidth]{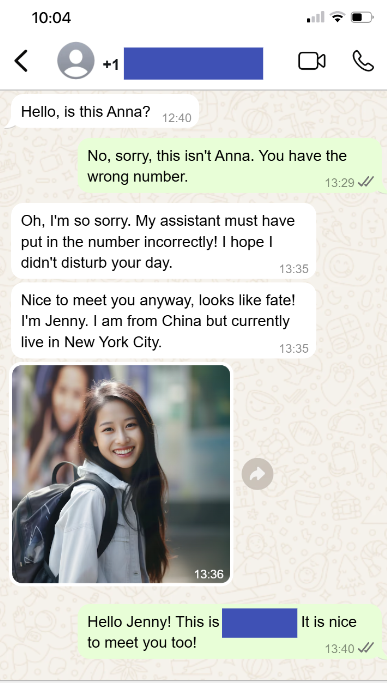}%\hfill
\caption{An example of how a Pig Butchering scam starts}
\label{fig:sample_chat}
\end{figure}

Pig butchering, also known as Shā Zhū Pán, represents a sophisticated and manipulative form of cyber-enabled financial fraud that has gained global traction in recent years~\cite{cross2024romance}. Originating in Southeast Asia~\cite{maras2024deconstructing}, this scam involves criminals establishing long-term relationships with victims through social media, dating apps, or messaging platforms~\cite{mnPartners24}. The term metaphorically describes the process where scammers ``fatten up" their victims—akin to fattening a pig before slaughter—by building trust and emotional connections over weeks or months before defrauding them of substantial amounts of money~\cite{FBI_IC3_2022}.
These scams typically begin with fraudsters reaching out to potential victims through social media, dating apps, or other online platforms~\cite{Sophos2023} (see Figure~\ref{fig:sample_chat} for the beginning of an actual pig-butchering scam~\footnote{This screenshot was provided by a participant in the study, but we have replaced the scammer's photograph with a stock image.}). Posing as successful investors, often in cryptocurrency, they build a relationship with the victim, sharing fake success stories and showing fake investment returns. %~\cite{Sophos2022}.
Once trust is established, victims are encouraged to invest small amounts of money, which appear to yield high returns. This lures them into investing larger sums, which the scammers eventually steal, leaving the victims financially and emotionally devastated. 
The impact of pig butchering scams is profound, affecting victims emotionally and financially. According to recent research~\cite{griffin2024crypto}, these scams have resulted in losses of nearly 75 billion dollars since 2020. The Securities and Exchange Commission (SEC) has recently also  filed suit against several entities for allegedly operating pig butchering scams~\cite{sec2024}.

Despite the growing prevalence of pig butchering scams, academic literature on the subject remains sparse. Existing research primarily focuses on broader categories of online fraud, such as romance scams~\cite{Rege2009}, investment fraud~\cite{Button2014}, social engineering tactics~\cite{Mitnick2002} and general strategies for prevention and detection of financial fraud~\cite{anderson2019financial}.  While these studies provide valuable insights into the mechanisms of online scams, there is a notable gap in the detailed understanding of pig-butchering scams, particularly regarding the intricate anatomy of these attacks, the specific strategies employed by fraudsters, and the unique psychological manipulation involved. 

\noindent \textbf{Motivations.}
%T. 
Pig-butchering scams differ from traditional  scams by combining prolonged emotional manipulation with sophisticated fraudulent investment schemes: unlike traditional romance scams, where scammers quickly request money often citing personal emergencies~\cite{Rege2009, whitty2015anatomy}, pig-butchering scammers invest months building trust and relationships and introduce victims to fake investment platforms. Unlike typical cryptocurrency scams that aim for quick financial gain without personal connections~\cite{xia2020characterizing, mirtaheri2021identifying}, pig-butchering involves deep emotional grooming. Pig-butchering is more organized and sophisticated as compared to traditional crypto scams~\cite{button2017cyber}.
Current reports offer situational awareness and anecdotal evidence, but lack the depth of systematic qualitative analysis from the victims' perspectives. In this paper, we present the first study on pig-butchering that draws from first-hand accounts from victims. We conduct in-depth semi-structured interviews with $N=26$ victims of pig-butchering so as to understand their experiences, strategies used by scammers, mental models and coping mechanisms. We seek to answer the following research questions:
\begin{list}{}{}
    \item \textit{\textbf{RQ1.}} What are the common characteristics and patterns in pig-butchering scams? 

    \item \textit{\textbf{RQ2.}} What strategies, tools and techniques are adapted by fraudsters in the various stages of pig-butchering scams?

    \item \textit{\textbf{RQ3.}} What psychological and financial impacts do pig-butchering scams have on victims?
    
\end{list}

\noindent \textbf{Contributions.}
In this paper, we offer the first end-to-end analysis of pig-butchering scams from a user perspective. Our findings illuminate the complex strategies employed by scammers and the multifaceted consequences faced by victims. We summarize our main contributions as follows:
\begin{itemize}
    \item We present a comprehensive anatomy of pig-butchering scams, breaking down the lifecycle of the scam into seven distinct stages, from initial contact through financial exploitation, and potentially onto an extended ``encore" scam.

    \item Through in-depth interviews with scam victims, we provide a detailed understanding of the emotional and psychological tolls, including shame, trust issues, and mental health impacts, offering qualitative insights into the broader consequences of these scams beyond financial loss.

    \item We find that victims often hesitate to report their experiences due to feelings of shame and fear of being ridiculed, which is further worsened by the stigmatizing term ``pig-butchering". Additionally, we find that for those who did report the scam, experiences were often frustrating as law enforcement was generally unable to provide meaningful help.

    \item Based on our analysis of scam patterns, we suggest intervention points for social media and financial platforms to detect and mitigate scams, particularly through identifying unsolicited contact patterns and verifying suspicious accounts or platforms.
\end{itemize}

\noindent \textbf{Paper Organization.}
The rest of the paper is organized as follows. Section~\ref{sec:related} reviews existing literature on online scams, particularly romance and investment fraud, to contextualize the unique features of pig-butchering scams. In Section~\ref{sec:methodology}, we describe our research design, including participant recruitment, interview procedures, and qualitative coding methods used to analyze the victim accounts. We then present an in-depth analysis of the stages and tactics within pig-butchering scams, detailing how scammers progressively manipulate victims in Section~\ref{sec:phases}. Section~\ref{sec:impact} examines the emotional and financial consequences of pig-butchering scams on victims. Finally, in Section~\ref{sec:discussion}, we synthesize our findings, offering implications for policy, social media and financial platform interventions, and future research.

\section{Related Work}
\label{sec:related}

\subsection{Cryptocurrency Scams}
Prior work has paid significant attention to characterizing and detecting cryptocurrency scams. As digital currencies become more mainstream, they have been increasingly exploited in these fraudulent schemes. 
Scammers often use cryptocurrency to facilitate transactions, making it more challenging for victims to recover lost funds~\cite{alekseenko2023, scharfman2024,agarwal2023}; they leverage the perceived profitability of crypto investments to entice victims. As described later in the paper, pig-butchering scams rely on the allure of high-return crypto investments to manipulate victims into repeatedly transferring large sums, believing they are engaging in lucrative trading activities.
Bartoletti et al~\cite{bartoletti2021cryptocurrency} developed a comprehensive taxonomy of cryptocurrency scams and proposed methods to detect and classify them effectively. Other work has studied more specific kinds of crypto scams in detail.
For example, Liebau and Schueffel~\cite{liebau2019cryptocurrencies}) investigate the prevalence of scams within Initial Coin Offerings (ICOs), suggesting that a substantial portion of these offerings could be fraudulent, potentially explaining the volatility in the cryptocurrency market. Similarly, Xia et al~\cite{xia2020characterizing} characterized cryptocurrency exchange scams, revealing significant financial losses due to scam domains and fake apps designed to deceive users. Li et al~\cite{li2023nothing} conducted an extensive study on giveaway scams, where attackers promise unrealistic returns to lure victims into sending funds. This study uncovered the large-scale nature of these scams, involving thousands of fraudulent websites and millions of dollars in losses. 

\subsection{Pig-Butchering Scams}
Pig-butchering scams, also known as \textit{Sha Zhu Pan}, are a form of online fraud where perpetrators cultivate long-term relationships with victims to defraud them financially~\cite{burton2024, ma2024}. This method relies heavily on social engineering techniques, where trust is cultivated over time, making victims more susceptible to manipulation~\cite{jiang2023}.
Prior work~\cite{jiang2023emotional} has highlighted the psychological aspects of these scams, noting that victims often exhibit low self-esteem, particularly in interpersonal-social dimensions, which scammers exploit to establish emotional trust before executing the fraud. \newadded{Complementing this work, Wang and Zhou~\cite{wang2023schemes} provide insights into the scammers' persuasive techniques. While they offer a valuable conceptual framing of pig-butchering scams through a three-stage model, our work substantially expands this foundation through empirical depth and granularity. They rely on publicly available reports to outline scammer strategies, whereas our study draws from 26 in-depth, first-hand interviews with scam victims, capturing the evolving tactics of perpetrators in real-world cases which go beyond their model.}
Legal responses to pig-butchering scams have been varied, particularly in regions like China, where the phenomenon has gained notoriety~\cite{wang2024}. From a legal standpoint, Tan~\cite{tan2023network} has found that the difficulty in identifying accomplices and collecting evidence complicates the prosecution of these cybercrimes, limiting the effectiveness of law enforcement. %The Chinese government has initiated campaigns aimed at curbing these scams, yet the problem persists and has expanded beyond national borders, affecting victims in Southeast Asia, North America, and beyond.

\subsection{Users \& Scams}
Studying scams through the lens of user experience provides valuable insights into the scam anatomy as well as effective prevention and intervention strategies.  
Miramirkhani et al~\cite{miramirkhani2016dial} conducted a large-scale analysis of technical support scams, which revealed that these scams exploit users' lack of technical knowledge, often through malvertising and fake tech support websites. Chen et al~\cite{chen2017securing} explored the factors leading to scam victimization, finding that routine internet activities and low self-control significantly increase the risk of being scammed. Research on online romance scams~\cite{sorell2019online, coluccia2020online} shows that victims often experience a double trauma of financial loss and emotional betrayal, which can lead to long-term psychological harm.
Understanding security experiences and behaviors of at-risk user groups has also been a focal point of research, showing how certain unique contextual factors amplify the risk of cybercrime for certain demographics~\cite{bellini2024sok, warford2022sok, frik2019privacy, ashraf2023stalking}. Herbert et al~\cite{herbert2024digital} compare and contrast cybercrime experiences between various at-risk user groups. Qualitative research has proven to be valuable tools in S\&P research~\cite{pfeffer2022replication, simoiu2019told, rostami2022being, faklaris2019self}.
The unique nature of pig butchering scams combining elements of romance, investment fraud, and advanced cyber tactics necessitates a focused analysis to understand its lifecycle fully. To this end, we aim to extend prior qualitative research on online scams to discover the specific strategies, psychological manipulation, and long-term grooming tactics that define pig-butchering.
\section{Methodology}
\label{sec:methodology}
We conducted in-depth semi-structured interviews with $N=26$ victims of pig-butchering. In this section, we describe our participant recruitment procedures, interview structure and analysis methodology.

\subsection{Participants}
\noindent \textbf{Recruitment. }
We recruited participants using social media outreach. We created a flyer and an accompanying text that introduced briefly the researchers, described our research and the nature of the study in short, and asked interested participants to fill out a screening survey. This flyer was advertised on X and through mailing lists. Additionally, we posted this flyer on Reddit in communities dedicated to discussing scams. Prior work has used Reddit extensively for user studies related to computer science~\cite{proferes2021studying}, and even cybersecurity and cryptocurrencies specifically~\cite{bouma2024honestly, floyd2021understanding}. A sample of our recruitment post in the form of a Reddit post has been provided in the Appendix.

\noindent \textbf{Screening.} Interested participants were sent a screening survey to screen for eligibility. We required that participants be at least 21 years of age, and able to converse fluently in English. We described pig-butchering in short and asked participants to attest that they had experienced it in some form. In order to filter fraudulent participants, the survey required participants to agree to showing us evidence of pig butchering and the experiences they describe during the subsequent interview. We excluded participants that did not meet these requirements. Out of 64 responses, we found 26 to be eligible. 

\subsection{Interview Design}
Our interviews were conducted online via Zoom. Considering the sensitive nature of the subject, we did not record the interviews. Instead, the interviewing researcher took detailed notes during the interviews, with participant consent.

\noindent \textbf{Overview. }
We first established rapport and trust with the participants by clearly stating the identity of the researchers and the purpose of the study. Participants were informed about the confidentiality of their responses, how the data collected would be used, and reported. Additionally, participants were reminded of their rights to decline to answer any questions or to terminate the interview at any point.

\noindent \textbf{Demographics. }
In this part of the interview, we collected basic demographic information to understand the background of each participant better. Questions were designed to gather age, gender identification, geographic location, highest educational attainment, field of education, current employment status, and marital status. 

\noindent \textbf{Scam Anatomy. }
We asked participants to describe their first encounter with the scam, how the scammer initiated contact, the platforms used for communication, and the nature of their interaction with the scammer. The objective was to understand the tactics used by scammers to gain trust and persuade victims to invest money, as well as to gauge the duration over which the scam was realized by the victim.

\noindent \textbf{Impact Assessment. }
The focus of this section was to assess the emotional and financial repercussions experienced by the participants due to the scam. Participants were asked about the impact on their emotional well-being and financial stability, the emotions felt upon realizing they had been scammed, and the total financial loss incurred. This part also explored whether any of the lost money was recovered, providing insights into the aftermath of the scamming experience.

\noindent \textbf{Support Systems. }
In this part, we investigated the support systems that participants turned to following the scam. Questions were directed to understand if participants sought help from law enforcement or support groups, and their satisfaction with the support received. The aim was to evaluate the effectiveness of existing support structures for scam victims and to identify areas needing improvement.

\noindent \textbf{Prevention. }
This section was designed to gather insights on hindsight awareness and preventive measures. Participants were asked about signs that could have alerted them to the scam earlier, their views on effective prevention strategies, and the role they believe social media platforms and financial institutions should play in protecting users.

\noindent \textbf{Conclusion. }
In this final section, we provided participants with an opportunity to add any further comments or share additional experiences that were not covered. We then thanked them for their participation, and ended the study. \\

\noindent Interviews were 38 -- 55 minutes long  ($\mu=44)$). Participants were compensated with a \$15 gift card in exchange for their time. The entire interview questionnaire is available in the Appendix. 

\subsection{Analysis}
Our interview was designed in a mixed-methods manner; questions we posed were a mix of quantitative and qualitative. For the quantitative questions, we coded participant responses based on well-known codes for the question (such as age brackets, academic qualifications, nationality). For the open-ended questions, we used open coding~\cite{braun2006using} inspired by grounded theory~\cite{glaser1967discovery, corbin2015basics}. Two coders were involve in the coding and analysis. Applying open-coding across each interview transcript, the two coders independently generated a set of codes. After completing the initial coding, the two coders met to compare and discuss their codes, identify areas of agreement and disagreement and work through any discrepancies. This resulted in generating of a common code-book. Another round of coding was done on the entire data using the shared codes. After obtaining a high inter-rating agreement (cohen's kappa $\kappa > 0.8$), codes were clustered together to form key themes.

\subsection{Ethical Considerations}
Throughout the study, we followed the best practices recommended for human subjects research according to the Menlo Report~\cite{kenneally2012menlo}. Participants were treated with dignity and respect throughout the study. Our study protocol was reviewed and approved by the Institutional Review Board (IRB) at our institution. We took several measures to ensure that our study posed minimal risk to participants.

\noindent \textbf{Informed Consent.} Prior to being enrolled in the study, we clearly informed participants of the purpose of the study, how the information they provide would be used and the nature of the questions asked. Participants were informed that they would never be identified by name, but may be quoted using anonymous identifiers. Additionally, we warned participants that recollections of being scammed may be traumatic or triggering, and that this was a risk in the study.

\noindent \textbf{Minimally Invasive Methods.} Our study methodology involved qualitative interviews, which are considered to be a minimally invasive research instrument~\cite{britten1995qualitative}. Considering the sensitive nature of the issue, we carefully structured our questions in a manner that did not ``victim-blame" participants for falling prey to scams. During the interview, we informed participants that they had the right to decline answering any question as well as the terminate the study at any point. 

\noindent \textbf{Privacy.} In order to protect the privacy and anonymity of subjects, we did not record or transcribe the interview. The interviewing researcher took detailed notes during the interview. Additionally, the identifiable information collected (email) was used solely to deliver compensation; this was not linked to interview responses in any way.

\subsection{Demographics}
Our study consisted of 26 participants, ranging in age from 27 to 68 years, with a mean age of 48 years. The majority of participants were male (80.8\%, $n=21$), with females comprising 19.2\% ($n=5$). Most participants held a Bachelor's degree (65.4\%, $n=17$), while others had attained a Master's degree (11.5\%, $n=3$), a PhD (7.7\%, $n=2$), or a high school diploma (11.5\%, $n=3$). Participants were primarily employed full-time (57.7\%, $n=15$), with others either retired (26.9\%, $n=7$) or employed part-time (12\%, $n=3$). The participants were geographically diverse, with the majority residing in the United States (53.8\%, $n=14$), followed by the United Kingdom (19.2\%, $n=5$), Canada (11.5\%, $n=3$), Germany (7.7\%, $n=2$), and Australia (7.7\%, $n=2$). The educational and professional backgrounds of participants spanned various fields, including Engineering, Finance, Communications, and Liberal Arts.

%Participants' annual income ranged from \$20,000 to \$147,000, with a mean income of approximately \$61,000. The amount of money lost due to the pig-butchering scams ranged from \$5,000 to \$160,000, with an average loss of approximately \$29,000.

\section{Lifecycle of Pig-Butchering Scams}
\label{sec:phases}

Pig-butchering is a complex and multifaceted fraud that unfolds in a series of carefully orchestrated steps, each designed to progressively manipulate the victim on emotional, psychological, and financial levels. The scam begins with the scammer making initial contact and establishing an emotional bond, then luring the victim into fraudulent investments under the guise of a promising opportunity. The scammer gradually escalates the stakes by fostering the victim's trust and dependency, pressuring them to make larger investments, and ultimately disappearing with the funds. In some cases, scammers reappear in an attempt to extract additional value from their victims. In this section, we provide a detailed breakdown of each phase, illustrating how these scams operate and the sophisticated tactics scammers employ to deceive their victims.

\subsection{The Lure}
This is the initial step in the pig-butchering scam, during which the scammer establishes contact with the intended victim. This phase is characterized by the creation of a carefully crafted and seemingly authentic persona. Victims are lured through social media platforms, dating applications, or unsolicited messages on popular messaging services. 

\noindent \textbf{Cold Texts.}
A significant proportion of participants reported that the initial contact with the scammer was through unsolicited messages via SMS, WhatsApp, or other messaging applications, where the scammer pretended to have entered an incorrect number. The first message of the form \textit{``Hello, is this} X\textit{"}, where \textit{X} was a random name. After being corrected by the victim, they would apologize but still introduce themselves and ask to continue the conversation. 

\noindent \textbf{Dating Apps.}
6 participants were approached on dating platforms (Tinder, Bumble, and Hinge). Scammers created attractive profiles, often featuring professionally taken photographs and bios highlighting shared interests with potential victims. Once a match was made, the scammers initiated conversations that quickly became personal, and attempted to move them off-platform on a messaging service like WhatsApp or Telegram.

\noindent \textbf{Social Media.}
A smaller segment encountered scammers through social media platforms (Instagram and LinkedIn). Scammers would send connection requests or follow the victims, often engaging by commenting on public posts, and initiate conversation via personal messages. Scammer profiles were carefully crafted and not obviously fake; \textit{P4} recounts: 
\begin{center}
    \textit{``...his LinkedIn profile was well-written, mentioned his projects and work experience in fair detail, and also had testimonials. He also had connections in the thousands.}
\end{center}

\begin{mdframed}[style=insightstyle]
Pig-Butchering scams begin with the scammer establishing contact with the victim through various online platforms. Scammers create convincing personas and use unsolicited messages or interactions on dating apps and social media to engage the target. This phase is characterized by the  construction of trust through seemingly incidental or calculated interactions.
\end{mdframed}

\subsection{The Bond}
Following the initial contact, the scam progresses into the \textit{Bond} phase, wherein the scammer attempts to build a deep emotional connection with the victim. They engage the victim in daily conversations and make a concerted effort to be involved in their lives. The scammer may share what seem to be intimate and personal details about their own life, fostering a sense of mutual trust and emotional closeness. The underlying goal during this phase is to make the victim feel that they are part of a genuine, caring relationship. This phase often spans several months, where the scammer patiently builds a relationship with the victim.

\noindent \textbf{Nature of Relationships. }
Scammers cultivated either business or romantic relationships with victims. The majority of participants developed what they believed to be romantic relationships with the scammers. These relationships were characterized by frequent communication, expressions of affection, and discussions about future plans together. The remaining participants were enticed through the prospect of business partnerships or mentorship. Scammers presented themselves as successful entrepreneurs or seasoned professionals willing to share their knowledge.

\noindent \textbf{Rapport Building Techniques. }
Scammers invested significant time in building rapport, with participants noting daily communication over several weeks or months. They shared personal stories, discussed family and career aspirations and showed empathy towards the victims' experiences. Scammers frequently used mirroring as a manipulation technique; participants noted that the scammer agreed with them, or reacted positively to any view of theirs. In the words of \textit{P8}:
\begin{center}
    \textit{``He agreed with my viewpoints on almost everything, from politics to lifestyle choices. It was like a perfect fit; we were on the same wavelength."}
\end{center}
Scammers also gradually disclosed personal information to build trust, sometimes sharing fabricated hardships or vulnerabilities, such as a challenging past or experiences with debt.

\noindent \textbf{Trust Building Techniques. }
Some scammers went even a step further in order to build more trust with victims. Participants reported that the scammer sent frequent pictures of themselves, and their surroundings. 7 participants reported having audio calls, and 3 participants video calls with the scammers; they were able to verify that the pictures sent matched with the person they spoke to on the video call. According to \textit{P2}:
\begin{center}
    \textit{``When I spoke to her on a video call, it was the same person from the photos. She was even wearing the dress that matched a photo she had sent earlier in the day."}
\end{center}

\noindent \textbf{Social Media. }
Scammers used well-designed social media profiles to keep up the appearance of a genuine personality, and to back up some of the claims they made to victims. Most participants (77\%) reported having connected with the scammer on at least one other social media platform: Facebook, Instagram, or X. Participants reported that the profiles were consistent with the details that the scammer shared, and that this was an important factor in them believing that the scammer was in fact a genuine person. For example, \textit{P18} shared:
\begin{center}
    \textit{``She said she liked dogs, and there was a picture of her with her dog on her Instagram. She said she went to [university] and her LinkedIn profile did say that too. It seemed legitimate, there was no reason to think that it could be fake -- if she could be a scammer, so could any of my actual friends."}
\end{center}
\noindent We requested participants to share links or screenshots of associated social media profiles. In all, we were able to collect $53$ profiles from various sites like Facebook, X, and LinkedIn. However, on following the links, or searching for the profiles based on the screenshot, we were unable to find even one. Our suspicion is that scammers churned profiles, and dumped an identity/social media profile after a successful attack. Additionally, because considerable time has passed since the original scam, the profiles may also have been taken down by platforms. 

\noindent \textbf{Duration. }
The bond phase showed that scammers had tremendous patience; this phase lasted anywhere from 3 to 11 months before the scammer moved on to the next stage of the scam. Scammers spent a significant amount of time bonding with the victim, to cultivate a sense of trust. According to most participants, this led them to letting their guard down and dispel any suspicions about this being a scam; \textit{P5} commented: 
\begin{center}
    \textit{``She did not push me to invest, or ask for money. She seemed genuinely interested in me and we spoke for nearly 6 months before she even brought up investments; it all seemed so real and organic."}
\end{center}

\begin{mdframed}[style=insightstyle]
Scammers deepen the relationship by fostering emotional trust with the victim through regular, personal interactions. They often simulate a romantic or business relationship that feels genuine to the victim. Over time, the victim becomes emotionally dependent, making them more susceptible to manipulation. The scammers' goal is to establish a strong emotional connection that will facilitate later exploitation.
\end{mdframed}
\subsection{The Bait}
Once trust has been firmly established, the scammer proceeds to the \textit{Bait} phase, where they introduce the fraudulent opportunity that forms the crux of the pig-butchering scam.

\noindent \textbf{Introducing Investments. }
Scammers typically introduce cryptocurrency investment into the conversation in a casual manner, referring to it as an extra source of income or even a hobby. They may claim to have insider knowledge or expertise in crypto trading, or an investment consultant who manages their portfolio. The bait is framed as an exclusive opportunity, often portrayed as an investment that is only available to a select few. 6 participants reported that the scammer described their investments in a manner aligned with the victims' circumstances, such as debts or future aspirations. In the words of \textit{P23}:
\begin{center}
    \textit{``When we were talking, I had mentioned that I have some student loan debt. She said that she also was in a lot of debt, but these investments have helped her pay off the debt, and start building her own wealth."}
\end{center}

\noindent \textbf{Providing Social Proof. }
Scammers provided fake screenshots or access to counterfeit investment platforms displaying impressive returns. Most participants were shown evidence of significant profits through fabricated screenshots. Scammers may also provide proof via testimonials; 5 participants reported that the scammer showed them emails or message screenshots confirming successful returns on investments. Further, some were also given emails and phone numbers of other supposed investors, so that they could act as references. Participant \textit{P20} says:
\begin{center}
    \textit{``He had a well-crafted LinkedIn profile, and several testimonials from other profiles that sung praises of his investment and market analytics skills. The other profiles looked real too; this was enough for me to trust him."}
\end{center}

\noindent \textbf{Leveraging FOMO. }
The scammer emphasizes the exclusivity and limited nature of the opportunity, thereby exploiting the victim's fear of missing out (FOMO). This approach is designed to make the victim feel privileged to be ``invited" into the scheme, which heightens the perceived legitimacy and desirability of the investment; most participants reported that the scammer presented the investment guidance as secret or highly specialized. At this point, the victim, already emotionally invested and trusting of the scammer, is primed to believe that this opportunity is both genuine and lucrative.
\begin{mdframed}[style=insightstyle]
After a bond has been established, the scammer introduces a fraudulent investment opportunity, typically framed as a low-risk, high-reward endeavor, such as cryptocurrency trading. The scammer uses fabricated success stories and social proof to persuade the victim of the legitimacy of the offer. This phase is designed to exploit the trust and rapport built in earlier stages, positioning the scammer as a trusted financial advisor or insider.
\end{mdframed}

\subsection{The Feed}
During the \textit{Feed} phase, the scammer transitions from enticing the victim to actively extracting financial contributions.

\noindent \textbf{Fraudulent Investment Platforms. }
Scammers provide access to sophisticated-looking but fraudulent investment platforms or applications, and asked victims to sign up for an account. All the participants reported creating accounts on such websites. 11 participants reported that the scammer helped them set up their accounts through step-by-step instructions; a few reported that there were even help articles or some form of customer support. These platforms were authentic-looking, and displayed real-time information, with  efforts taken to ensure that they resembled real websites. \textit{P22} says:
\begin{center}
    \textit{"[the site] was similar to what you would expect on an investment portfolio website; in fact, the prices of stocks and bitcoin also matched..."}
\end{center}
\noindent Similar to social media profiles, we attempted to analyze the fraudulent investment platforms. We were able to collect a list of $23$ apps/platforms from participants, but found that the websites were no longer active. Our hypothesis is that since it is trivial to purchase new domain names, scammers frequently change domains as an evasion tactic. 

\noindent \textbf{Initial Investments. }
The scammer typically begins by encouraging the victim to make a modest initial investment. Participants reported initial investments of \$20 -- \$125. This allowed the victim to test the waters without feeling at great risk, and it also provides an opportunity for the scammer to build credibility. To maintain the illusion of legitimacy, the scammer may manipulate fake trading platforms or falsify investment returns, making it appear as if the initial investment has yielded positive gains; all the participants saw significant gains in their initial investment (2x -- 5x). 3 participants also reported being allowed to withdraw some amount, further convincing them of the credibility of the investments. As \textit{P1} says:
\begin{center}
    \textit{"I was skeptical, and put in only \$25, but it soon became \$100, just as she had predicted! I was able to withdraw it in the form of a gift card too -- the investment strategy seemed legitimate."}
\end{center}

\noindent \textbf{Escalating Investments. }
Scammers applied subtle pressure to reinvest profits rather than withdraw them. Most participants were encouraged to roll over their gains into new investments, so as to earn even more. As victims grew more comfortable, scammers suggested larger investments to achieve greater profits. According to 13 participants, victims were encouraged to use their 'profits' as leverage for higher-tier investment opportunities. All participants reported increasing their investment amounts significantly (\$3,500 -- \$65,00) by putting in more money. Participants also saw tremendous gains on their investments on the platform (2x -- 12x).

\noindent \textbf{Manipulation Techniques. }
Scammers maintained regular communication, celebrating the victim's 'successes' and reinforcing the notion of smart investing; participants received congratulatory messages whenever their account showed any gains. Scammers often claimed to be investing alongside the victims, fostering a sense of shared endeavor; they manipulated the investment platform so as to create this illusion, even 'lending' money if needed. \textit{P14} recounts: 
\begin{center}
    \textit{"..there was this particular opportunity that had a \$1000 minimum, but I did not feel confident enough and had only \$500 in the account; she said she can front me the remaining amount (as it would be for our future together) -- and the amount really did arrive as a transfer to my account [on the platform]!"}
\end{center}
Further, according to some participants, scammers introduced complex investment strategies, using technical jargon to appear knowledgeable. Finally, all participants reported that scammers discouraged them from discussing investments with others, citing reasons like market confidentiality.
\begin{mdframed}[style=insightstyle]
    Scammers persuade victims to invest in fraudulent platforms that mimic legitimate financial services. Initial investments are typically small, with fabricated returns used to encourage further contributions. The scammer maintains ongoing communication, reinforcing the illusion of success and increasing the victim’s financial commitment by encouraging reinvestment of perceived gains.
\end{mdframed}

\subsection{The Squeeze}
In the \textit{Squeeze} phase, the scammer escalates the pressure on the victim to act with urgency. By this stage, the scammer has already secured the victim's trust and convinced them of the legitimacy of the investment scheme, making the victim highly susceptible to psychological manipulation. 

\noindent \textbf{Exclusive Deals. }
After continuing investments, all participants were presented with time-sensitive, high-return investment opportunities. Scammers claimed access to insider deals or initial coin offerings (ICOs) about to explode in value. These investments had minimum amounts, which were often large (in the tens of thousands). Most participants reported that they felt inclined to invest, because of the perceived returns so far.

\noindent \textbf{Pressure Tactics. }
To ensure the high investment, scammers increased the frequency and urgency of their messages. All participants reported receiving multiple messages or calls daily, emphasizing the need for immediate action. Scammers employed emotional manipulation, expressing disappointment or suggesting that the victim was missing out due to hesitation. Some participants, for whom the relationship was of a romantic nature) also reported frustruation on the scammer's part; as \textit{P10} put it: 
\begin{center}
    \textit{``[she] got extremely angry, and asked me whether I was actually serious about settling down with her, and if so, why I wasn't making an effort to put in this money."}
\end{center}

\noindent \textbf{Continuing Investments. }
Scammers encouraged victims to ``re-invest" their gains, and add more funds rather than attempt to withdraw them. Exclusive opportunities as described above keep on occurring till the victim invests. The squeeze serves to drain any remaining financial resources the victim may have access to, pushing them to borrow money or liquidate other assets if necessary. 9 participants reported that the scammer recommended using credit cards or applying for new credit cards so as to use credit to invest. 
The squeeze phase lasts for as long as the victim is willing to invest; participants reported continuing large investments and perceived profit for 1-7 months. By the end of this phase, the scammer aims to have extracted as much financial value as possible.

\begin{mdframed}[style=insightstyle]
    Once the victim is fully engaged, scammers escalate pressure by introducing time-sensitive, high-return investment opportunities, often requiring large additional investments. Emotional manipulation intensifies, with scammers leveraging the victim's trust and fear of missing out (FOMO) to secure further financial commitments. This phase aims to extract as much financial value as possible from the victim.
\end{mdframed}

\subsection{The Cut}
This phase marks the culmination of the pig-butchering scam, during which the scammer absconds with all of the victim's invested funds. The scammer transitions into the cut phase after they are convinced that they have squeezed the victims of the maximum amount possible. 

\noindent \textbf{Stalling Tactics. }
The cut phase typically begins when victims refuse to invest any more money, and attempt to withdraw funds. Scammers used several tactics to delay the withdrawal; most participants discovered that they could suddenly no longer access their investment accounts. 10 participants reported that the platforms displayed error messages or indicated that the account was under review. Some scammers went a step further; 16 participants reported that the platform asked for more money under the guise of withdrawal fees or taxes. 4 participants fell for this and ended up paying. \textit{P19} says:
\begin{center}
    \textit{``At the time, I had [perceived] \$250k in the account, and when I attempted to withdraw it, they said that 30\% had to be paid as taxes. When I asked them to deduct it from the account, they refused, claiming that the money was tied up in crypto and government regulations needed taxes to be paid directly. I even chatted with a support representative on the site." }
\end{center}

\noindent \textbf{Complete Disappearance. }
After either paying fees for withdrawal, or refusing to do so, all but one participants reported the scammer ceasing all communication. Scammers blocked the victims on chat application (WhatsApp / Telegram), as well as any social media platforms they were connected on. Scammer accounts were likely deleted or deactivated; participants searched for the profiles using other accounts, but could not find them. \textit{P22} says:
\begin{center}
    \textit{``I used my sister's Instagram to look for her, but I couldn't find her profile. I also tried messaging her on WhatsApp using my sister's number, but the app said that this number wasn't on WhatsApp."}
\end{center}
\begin{mdframed}[style=insightstyle]
    After squeezing the victim as much as possible, scammers abscond with the victim’s funds, typically after the victim attempts to withdraw their investments. Scammers employ stalling tactics, such as claiming technical issues or additional fees, to delay withdrawal. Once the victim realizes they have been defrauded, the scammer cuts all communication, disappearing from both the financial platform and social media.
\end{mdframed}

\subsection{The Encore}
Even after disappearing, scammers may attempt an \textit{Encore} by re-engaging the victim for additional scams. This phase leverages the victim's desperation to recover their lost funds and exploits the emotional turmoil resulting from the initial scam. After extracting the money, encore scammers block the victim and disappear.

\noindent \textbf{Impersonation Scams. }
Scammers frequently impersonate law enforcement officials to re-establish contact with victims. They pose as detectives, federal agents, or members of cybercrime units investigating the original scam. Most participants reported receiving official-looking communications, complete with badges, case numbers, and legal jargon. Some reported that the supposed agent requested personal information under the guise of needing it for the investigation. According to \textit{P12}:
\begin{center}
    \textit{``I received an email from someone claiming to be an FBI agent, stating that they had found my account details when they took down the platform, and that they were closing in on the scammers who took my money. They asked for my bank statements and ID to 'verify' my case, and I complied without thinking twice."}
\end{center}
Further, a few participants reported that these fraudulent officials informed them that their funds have been located but require payment of legal fees or taxes to release them. Overall, 15 participants were victims of impersonation scams, and ended up paying additional amounts of \$120 -- \$1700. 

\noindent \textbf{Recovery Scams. }
Scammers, sometimes the original perpetrators or new opportunists, contact victims offering services to retrieve lost funds. 11 participants reported being contacted by such recovery experts. They present themselves as recovery specialists, private investigators, or legal professionals with expertise in financial fraud cases . They typically request upfront fees for their services, citing costs for legal processes, court fees, or administrative expenses. Overall, 11 participants were approached with recovery scam offers. These scams can be highly sophisticated; \textit{P20} says:
\begin{center}
    \textit{``They had a website and a case management portal; there were many different professionals within the company; a case manager, a lawyer, a digital forensics expert...."}
\end{center}

\noindent \textbf{Legal Threats. }
Scammers may also employ legal threats to intimidate victims into making additional payments. They impersonate lawyers, debt collectors, or officials from regulatory agencies, claiming that the victim has breached contracts or violated laws through their previous investment activities. The threats typically include demands for immediate payment to avoid lawsuits, arrests, or damage to credit ratings. Scammers may threaten to report the victim to authorities for participating in illegal investment schemes or non-payment of taxes. \textit{P21} reported:
\begin{center}
    \textit{``I received an email from the platform saying that they were owed some fees from me, and if I do not comply, then they would be filing an official complaint with the IRS for tax evasion. "}
\end{center}
%Overall, 7 participants reported being contacted with legal threats; however, fortunately, none of them fell for it. 
\begin{mdframed}[style=insightstyle]
    In some cases, scammers re-engage victims by posing as law enforcement or recovery agents, offering assistance in recovering lost funds for a fee. This phase exploits the victim’s emotional vulnerability and desperation, further compounding their financial losses. 
\end{mdframed}

\section{Impact on Victims}
\label{sec:impact}
The consequences of pig butchering scams extend far beyond the immediate financial losses. Victims endure profound financial hardships, emotional turmoil, and often face challenges when seeking assistance from law enforcement. In this section, we describe the effects of the scam on vitims, and their experiences in reporting them. 

\subsection{Financial Impact}
The financial repercussions for victims of pig butchering scams are severe and long-lasting. All participants reported significant monetary losses, with amounts ranging from \$600 to as high as \$185,000. 

\noindent \textbf{Loss of Savings.} A significant number of victims experienced devastating losses to their personal savings. Many had invested substantial portions of their life savings, retirement funds, or college funds for their children, enticed by the promise of high returns. Financial strain affected victims' ability to meet daily living expenses, and resulted in long-term financial insecurity. Some reported struggling to pay for rent, medical care or other bills. 6 participants (who were retired or nearing retirement age) had used their retirement savings to invest; \textit{P22} says:
\begin{center}
    \textit{``I was planning to retire in three years, and I pulled out my 401(k) for a more comfortable retirement....but now I don't know if that will ever be possible."}
\end{center}

\noindent \textbf{Accumulation of Debt.} Many victims accrued substantial debt as a result of the scam. A few resorted to personal loans, borrowing from friends and family, and liquidated an existing asset to invest more heavily under the scammer's influence. In the words of \textit{P25}:
\begin{center}
    \textit{``...my investments seemed to be growing so rapidly, I took out a second mortgage on my home to maximize the investment, believing the returns would cover the payments easily."}
\end{center}
Further, participants reported accruing significant credit card debt. High-interest rates associated with credit cards further exacerbated their financial difficulties post-scam. Some even reported damaged credit scores due to missed payments and increased debt-to-income ratios. 

\noindent \textbf{Ripple Effect. }Finally, for 5 participants, financial losses affected their dependents and family members significantly. \textit{P16} recounts:
\begin{center}
    \textit{``My children's college funds were wiped out, I believed I was doing all this investing for a more stable future for them, and now I feel immense guilt for jeopardizing their future."}
\end{center}

\subsection{Emotional Impact}
The emotional toll on victims was profound, encompassing a range of negative feelings and psychological challenges. 

\noindent \textbf{Shame and Embarassment. }The most common feelings experienced by victims were shame and embarrassment, as reported by most participants. \textit{P3} expressed:
\begin{center}
    \textit{``I felt so ashamed for being deceived. I considered myself savvy, but I fell for it."}
\end{center}
Feelings of stupidity and self-blame were also common, as victims grappled with how they could have been manipulated. A few reported that they blamed themselves for falling for something that was obviously a scam. The stigma associated with being scammed heightened their sense of humiliation; \textit{P25} mentioned:
\begin{center}
    \textit{``I felt so stupid that I lost all this money to such a scam, I didn't tell my family or friends about it because I was ashamed of what happened."}
\end{center}

\noindent \textbf{Loss of Trust. }
Loss of trust was another critical emotional impact; betrayal by scammers severely impacted victims' ability to trust others. According to most participants, it was hard to trust anyone online, even if they were previously known to them, fearing that a scammer was using a fake profile or hijacked account. Participants who developed romantic relationships during the scam reported experiencing feelings akin to heartbreak; \textit{P26} says:
\begin{center}
    \textit{``It felt like I was betrayed by the love of my life; I could not stop thinking about her. I do not think I will ever find true love again, I cannot trust anyone this much again..."}
\end{center}

\noindent \textbf{Emotional Distress \& Mental Health.} The emotional distress caused by the scam had significant implications for victims' mental health. Many reported being clinically diagnosed with mental health conditions such as depression, anxiety, and even post-traumatic stress disorder (PTSD).
Paranoia and hyper-vigilance were also reported; some described themselves as being constantly on edge, fearing further exploitation. \textit{P6} mentioned:
\begin{center}
    \textit{``I was paranoid about every small thing, every email, text message and phone call, worried that someone was trying to scam me again."}
\end{center}

\subsection{Seeking Help from Law Enforcement}
Victims' experiences with law enforcement were mixed, often marked by frustration and disappointment.

\noindent \textbf{Reluctance to Report.}
A significant number of participants did not report the scam to authorities due to feelings of shame and embarrassment.  Feelings of shame and self-blame deterred many from coming forward. In the words of \textit{P7}:
\begin{center}
    \textit{``I was too humiliated to go to the police. I responded to a wrong number text, and she became my girlfriend; the police would laugh at me for falling for it." }
\end{center}
The stigma associated with being a victim of fraud contributed to a culture of silence; according to 7 participants, they did not want to come forward with their stories because they did not want anyone to know that they had been a victim. \textit{P8} said:
\begin{center}
    \textit{``I did not want to be known as the woman who foolishly fell for a scam...what if I had to appear in court or something? Everyone would know that I was one of those...my friends, family, co-workers would know." }
\end{center}

\noindent \textbf{Limited Assistance from Authorities.}
Those who did report the scam often found that law enforcement could do little to assist. Only 7 participants reported lodging a complaint with their local police station, and 3 with the FBI. Out of those who sought help from their local police, 6 were informed that, without formal contracts or identifiable perpetrators, pursuing the case was challenging. 2 were also told that the transactions weren't fraudulent; \textit{P21} mentioned:
\begin{center}
    \textit{``The police took my statement but said there wasn't much they could do since the scammers were likely overseas and used untraceable methods. Moreover, I had never signed a formal contract, nor did they steal as I transferred the money willingly."}
\end{center}
The complexity of the scams, involving international elements and sophisticated cyber tactics, posed significant barriers to investigation. 3 participants were informed by law enforcement agencies that as the scammers were probably outside the US, they lacked the resources or jurisdiction to pursue the perpetrators effectively.

\subsection{Pig Butchering Terminology}
The term "pig butchering" elicited strong emotional reactions from the victims interviewed in this study. Most participants expressed discomfort and offense at being associated with the term, feeling that it dehumanized their experiences and added insult to injury.\textit{P7} remarked:
\begin{center}
    \textit{``Being compared to a pig made me feel even more humiliated, and made me out to be an animal rather than a human. As if I was naive and deserving of what happened, and that my experience was to be expected."}
\end{center}
Most also felt that the term trivialized their ordeal, reducing a complex and traumatic experience to a crude metaphor. In the words of \textit{P11}:
\begin{center}
    \textit{``People laugh it off because of the term pig-butchering. They laugh at the victims who were pigs. What happened to me was real, I lost actual money, and the ability to trust people again -- it wasn't a joke." }
\end{center}
Further, participants also felt that the term was mocking, and that that was one of the contributing factors in them not reporting the scam. \textit{P13} stated:
\begin{center}
    \textit{``The label made it harder to talk about what happened. The name is degrading, and makes fun of the victim; I did not report it or speak about it people might mock me or ridicule me for falling prey to such a scam."}
\end{center}
\section{Discussion}
\label{sec:discussion}
\subsection{Key Takeaways}
\noindent \textbf{Complex Anatomy of Pig Butchering Scams.} Our study finds that pig butchering scams follow a highly structured and carefully orchestrated lifecycle, moving through defined stages that strategically manipulate the victim both emotionally and financially. Each phase, from initial contact through emotional bonding to escalating financial exploitation, plays a specific role in priming the victim. For instance, 76.9\% of participants reported that scammers built trust over extended periods—sometimes up to 11 months—before introducing investment opportunities. Additionally, scammers sent pictures, and even engaged in audio and video calls with victims. Because social media plays an important role in the lure and bond phases, effective intervention points exist at the platform level. For instance, platforms could develop automated detection systems to identify and block patterns of unsolicited messages and flag prolonged manipulative engagements, particularly those linked to unverified accounts. 

\noindent \textbf{Sophisticated Execution. }
We find that participants reported using sophisticated, realistic online platforms that simulated genuine financial services. These platforms displayed detailed data, including real-time stock prices, which made the scam appear more legitimate. Scammers further built credibility by showing participants fake screenshots of supposed profits; 84.6\% of participants received these fabricated results, reinforcing the scam's illusion of authenticity. Some victims (11.5\%) were even permitted to withdraw small amounts, which enhanced trust in the platform. These findings emphasize how scammers exploit the appearance of legitimacy to build credibility. For prevention, financial and technology sectors could collaborate to develop verification standards or security markers for investment platforms, helping users distinguish legitimate services from scams. User literacy programs focusing on digital security skills could also provide guidance on verifying websites, identifying phishing techniques, and understanding the risks of online investment, helping to mitigate the influence of realistic-looking fraudulent platforms.

\noindent \textbf{Emotional Manipulation. }
We find that emotional manipulation is central to pig butchering scams. Scammers frequently simulate emotional intimacy, as indicated by the fact that 61.5\% of participants believed they were in a romantic relationship with the scammer. Scammers used daily communication and mirroring tactics, with 34.6\% of participants noting that scammers seemed to agree on every topic discussed. The trust built through this constant reinforcement laid a foundation for subsequent manipulation. We also find that scammers often tap into victims' financial insecurities, such as debt, retirement concerns, or aspirations for financial independence. By aligning their pitches with victims' personal financial situations, scammers can frame their investment opportunities as solutions to specific financial challenges. This finding underscores the need for public awareness efforts that educate people about the risks of intense online relationships and the psychological techniques fraudsters use to foster dependency. 

\noindent \textbf{Financial \& Emotional Impact.}
Our study shows that the financial repercussions of pig butchering scams extend beyond the initial loss, often leading to severe debt accumulation, and even affecting victims' families. Victims in the study reported losses as high as \$185,000, with many turning to personal loans, credit cards, and asset liquidation to fund their ``investments". The emotional aftermath of pig butchering scams is also substantial. Victims reported a range of negative psychological effects, including shame, embarrassment, and self-blame. A striking 73\% of participants expressed feelings of humiliation, which deterred many from discussing their experiences with others or seeking help. Of particular note, 26.9\% of participants said that fear of judgment stopped them from reporting the scam to authorities.  %Financial institutions should consider implementing alerts for unusual transactions that could signal fraud, providing a layer of protection for at-risk customers.

\noindent \textbf{Secondary Scams.}
Disturbingly, we find that scammers often re-target victims through secondary scams after the initial fraud, exploiting victims’ desperation to recover their losses. Approximately 57.7\% of participants received follow-up contacts from individuals posing as law enforcement officers or recovery agents, who claimed they could help recover the stolen money for a fee. These secondary scams are often sophisticated; some participants were presented with “official” documents and identification, while 34.6\% were asked to pay upfront fees for “legal” or “tax” processing.
The encore phase compounds victims’ financial and emotional losses, demonstrating scammers' persistent exploitation of vulnerable individuals. 

\noindent \textbf{Stigmatizing Terminology.}
The term “pig butchering” has a stigmatizing effect on victims, exacerbating feelings of shame and reducing the likelihood of reporting. About 69.2\% of participants found the term offensive and dehumanizing, with some expressing that it trivialized their experience. This language can dissuade victims from seeking assistance, as they fear public ridicule and judgment, further isolating them and prolonging the impact of the scam.

\subsection{Implications}

\noindent \textbf{Social Media.}
We find that social media and messaging platforms are often the first points of contact for pig butchering scams, making them a critical point of intervention. Platforms should invest in AI-driven tools to detect unsolicited messages, especially those that follow known scam patterns, such as "wrong number" texts used to initiate contact. Implementing automated alerts for users when suspicious activity is detected can help individuals identify potential scams early on. Platforms could also verify accounts more rigorously to prevent scammers from creating fake profiles that appear credible to potential victims. Educating users on the red flags associated with scams should be part of social media companies' responsibilities; integrating brief educational prompts or scam prevention content into user feeds could raise awareness.

\noindent \textbf{Banks \& Financial Institutions.}
Financial institutions have a pivotal role in detecting and preventing pig butchering scams by monitoring transactions for signs of unusual activity. Banks should implement advanced transaction monitoring systems that flag high-risk transactions, particularly sudden large transfers to new accounts, which are common in these scams. These systems could work in tandem with AI-driven fraud detection tools that analyze spending patterns and identify anomalous transactions indicative of scam behavior. Financial institutions should also provide customers with warnings before they proceed with large, out-of-character transfers, potentially triggering an automated advisory about pig butchering scams. In addition, banks could offer targeted financial counseling for individuals affected by scams, assisting them in managing debt and loss recovery. Institutions could collaborate with law enforcement to improve detection and reporting processes, creating a feedback loop that helps combat scam operations.

\noindent \textbf{Law Enforcement.}
For law enforcement, the complexity and anonymity of pig butchering scams pose a significant challenge, as victims reported high levels of frustration with the response they received. Strengthening cybercrime units with specialized training and resources dedicated to tackling financial scams is essential to improve response effectiveness. Law enforcement agencies could benefit from establishing victim-focused units trained in cyber psychology and financial fraud to ensure sensitive handling of such cases. Improved digital forensic capabilities would enable agencies to track scammers more effectively, even across borders. Additionally, law enforcement agencies should coordinate with banks and financial institutions to monitor unusual transactions that could signal fraud. Cross-agency collaboration between cybercrime units and financial regulators could help streamline response times and increase scam intervention success rates. %Given the stigma victims feel, law enforcement should also create accessible, stigma-free channels for victims to report scams anonymously.

\noindent \textbf{Government Policy \& International Cooperation.}
The cross-border nature of pig butchering scams, with perpetrators frequently operating from regions with limited extradition agreements, underscores the need for governments to prioritize international cooperation in cybercrime investigations. These scams are often highly organized, requiring governments to strengthen cross-border frameworks that facilitate information sharing, investigation, and prosecution. Current anti-fraud laws are not uniformly enforceable internationally, which limits the ability to pursue justice for victims. Governments could support dedicated cybercrime units and establish clear frameworks for joint operations with other countries to improve response rates. Awareness campaigns run by governments can also play a vital role in educating the public on these scams, emphasizing their sophisticated emotional and psychological aspects. %Through a concerted policy focus, governments can better protect citizens from becoming victims of international fraud schemes.

\noindent \textbf{Rebranding Pig-Butchering.}
Replacing the term "pig-butchering" with more neutral, non-stigmatizing terminology could encourage victims to come forward and report their experiences. Adopting terminology that accurately describes the manipulative tactics without demeaning victims can reduce the stigma associated with reporting and discussing the scam. For example, terms like \textit{Grooming Investment Scam}, \textit{Long-Con Investment Fraud} or \textit{Groom-and-Swindle Scheme} capture the essence of the scam in a non-derogatory way. \newadded{We still use the term ``pig-butchering" because at the time of writing, it was the most widely recognized term in public discourse, law enforcement advisories, and emerging literature. By first anchoring the discussion using the recognizable term, and then problematizing it with evidence from victim narratives, we believe we can effectively advocate for re-framing in the terminology.}

\subsection{Limitations}
While this research provides essential insights into the mechanics and impacts of pig-butchering scams, some limitations warrant acknowledgment. The sample size of 26 victims, though yielding valuable qualitative depth, may not fully capture the diverse experiences across broader demographics and regions impacted by these scams. As the findings rely on retrospective and self-reported accounts, recall bias and reluctance may affect accuracy. Despite these limitations, this research offers a crucial foundation for understanding pig-butchering scams and highlights pathways for future studies to build upon and broaden these findings.
\section{Conclusion}

Our study provides the first comprehensive examination of pig-butchering scams through in-depth interviews with victims. Our findings reveal that these scams are meticulously orchestrated, combining elements of social engineering, emotional manipulation, and fraudulent investment schemes to exploit victims over extended periods. By examining the lifecycle of pig-butchering scams, we identify distinct phases, from initial contact to emotional bonding, and eventual financial exploitation, each designed to maximize the scam’s impact. We also find that the scammers' use of sophisticated techniques, including fake financial platforms and personalized engagement, makes these scams particularly difficult to detect and combat. We find that the impact of these scams on victims is significant; causing significant financial losses, emotional distress, and long-term impacts on their trust in others. Our research highlights the urgent need for targeted interventions by social media platforms, financial institutions, and law enforcement to curb these scams.

\bibliographystyle{plain}
\bibliography{refs}

\appendices
\section{Recruitment Materials}
\subsection{Flyers}
\noindent As described in Section~\ref{sec:methodology}, we recruited participants via Twitter(X) and Reddit. Below, we provide the recruitment flyer posted on Reddit. The flyer for X contained the same information, but in a more condensed form and posted as an image. \\
\hrule
\begin{center}
    \textbf{\large Research Study on Pig-Butchering Scams}
\end{center}
\noindent Hello, we are researchers from UC Davis. We are conducting a research study on pig butchering scams, seeking participants who are willing to share their experiences through an informational interview. We will anonymize any insights collected from the interviews, and you will never be identified by your name.

\noindent If you are 21 years or older and have been a victim or a near-victim of a pig butchering scam, your insights could be incredibly valuable for this research. We will provide you a \$15 Amazon Gift Card in exchange for your participation. Your insights will be really invaluable to systematically study these scams, understand their MO, spread awareness and prevent them from being successful.

\noindent Please fill out this form if you are interested: \texttt{\textcolor{blue}{LINK TO FORM}}

\noindent We will review your answers and reach out if you qualify for the study. Note that we will use the email address only to contact you; once we have screened you for eligibility, we will not link it to your interview.

\noindent This study is approved by the Institutional Review Board (IRB) at UC Davis. The proof of approval can be found here: \url{https://tinyurl.com/pbirbapproval}

\noindent If you have any questions, please feel free to contact the lead researcher: rvoak@ucdavis.edu.

\subsection{Screening Survey}
\noindent Interested participants were directed to complete a screening survey. Below, we provide the questions asked in this survey.\\
\hrule
\begin{center}
    \textbf{\large Screening Questionnaire: Pig Butchering}
\end{center}
\noindent This is a screening questionnaire for a research study being conducted by researchers from UC Davis. The goal is to gain a deeper understanding of pig-butchering scams, discover strategies employed by scammers and identify potential mitigation measures.

\noindent The actual study will be in the form of an online interview lasting ~45 minutes. We will not ask for your name or any other personally identifiable information. You will be assigned an anonymous identifier, and any quotes will be attributed to this identifier only. More information about the study can be found here: \url{https://tinyurl.com/pbinfosheet}.

\noindent Please fill out the below questionnaire to see if you qualify. Qualifying participants will be compensated with a \$15 Amazon Gift Card in exchange for their time.

\noindent This study is approved by the Institutional Review Board (IRB) at UC Davis. If you have any questions, please feel free to contact the lead researcher: Rajvardhan Oak (rvoak@ucdavis.edu).\\

\noindent \textbf{\large Email \& Age Confirmation}
\begin{enumerate}
    \item What is your name? This will be used only to contact you and schedule an interview if you qualify. We will not link this with your interview responses.
    \item What is your email address? 
This will be used only to contact you and schedule an interview if you qualify. We will not link this with your interview responses. 
    \item Are you 21 years or older? \textit{Yes / No}
\end{enumerate}

\noindent \textbf{\large Experience with Pig Butchering}\\
\noindent \textit{A pig butchering scam is a type of financial fraud where scammers build a fake relationship with a victim, often posing as a romantic interest or a business partner, to gain their trust. A pig butchering scam typically begins with the scammer establishing a relationship with the victim. This could be through dating apps, social media, or other online platforms. Once the scammer has established this trust, they convince the victim to invest money in a non-existent or fraudulent scheme.}
\begin{enumerate}
    \item Have you been a victim of a pig-butchering scam? Being a “victim” means that you lost some amount of money as a result of the scam. \textit{Yes / No}
    \item Have you been a near-victim of a pig-butchering scam? A near-victim means that the scam was successful to some extent, but you did not lose any money. \textit{Yes / No / Maybe}
    \item If you selected "Maybe" above, please describe your experience in short.
    \item If required, will you be able to show us proof of the scam (such as chats with the scammer)? We will not collect these chats or use them in our research; these will simply serve as a confirmation of your eligibility. \textit{Yes / No} 
\end{enumerate}

\section{Consent Form \& Information Sheet}
\noindent The following information sheet (which also served as a consent form) was provided to participants before the interview began.\\
\hrule
\begin{center}
    \textbf{\large INFORMATION SHEET}
\end{center}
\subsection*{Overview}
\noindent This is a study being conducted by researchers from UC Davis. The goal is to gain a deeper understanding of pig-butchering scams, discover strategies employed by scammers and identify potential mitigation measures. The study protocol is approved by the Institutional Review Board(s) at UC Davis, and you may view the approval letter here: \url{https://tinyurl.com/pbirbapproval}.

\subsection*{Nature of Study}
\noindent The study consists of an informational qualitative interview. You will be posed a series of semi-structured questions designed to understand your experiences with pig-butchering scams, including but not limited to how the scam began, the nature of your relationship with the scammer, and information about any investments you made. The interview will be conducted online via Zoom and will not be recorded. Participants will be provided an Amazon Gift Card worth \$15 after successful completion of the interview. 

\subsection*{Data Collection \& Use}
\noindent We will collect your email address as part of the study, as it will be required in order to contact you regarding potential compensation and delivery of e-gift cards. The email will not be stored, or used as a linking key. 
Any quotes that we use will be only attributed to a pseudonym (P1, P2, etc) and will not identify you by name. 
We will retain the pseudonymized data indefinitely, and may make it available to other researchers if needed. 

\subsection*{Potential Risks}
\noindent Because we will not be storing personally identifiable information, participation in this study does not involve a risk of your anonymity being compromised. 
Questions we ask will involve your past experiences with scams and the scammer, and it may be a traumatic experience to relive them.

\subsection*{Participant Rights}
\noindent Participation in the study is voluntary. 
Providing demographic information is voluntary. If you do not wish to provide the information, please mark “Prefer not to disclose” as the answer. 
You may decline to answer any questions that you deem inappropriate, or terminate your participation at any time.

\subsection*{Grievances}
\noindent If you have any questions or concerns about the study, or the compensation policy, you may contact the researchers directly:
\begin{itemize}
    \item Rajvardhan Oak (rvoak@ucdavis.edu)
    \item Zubair Shafiq (zubair@ucdavis.edu)
\end{itemize}

\hrule

\section{Interview Questionnaire}
\noindent Below, we provide the script we used during interviews. Note that as this is qualitative research, we made minor adjustments as needed during the actual interviews, and probed deeper with additional questions where necessary. 

\subsection*{Overview}
\noindent Hello! Thank you for joining us today. My name is Raj, and I am part of the research team at UC Davis. I want to begin by thanking you for your willingness to share your experiences with us. Your insights will play an essential role in helping us understand the mechanisms of scams like the one you experienced and how we can better protect others. Before we begin, I want to assure you that everything you share in this interview will remain confidential. We will not be recording this conversation. Instead, I will be taking detailed notes to capture your insights, with your consent. The purpose of this interview is to learn more about your experiences. Our goal is to understand how scams like pig-butchering operate and to identify ways to improve support for those who may be affected. All data collected from our conversation will be used for research purposes only. In our reports, no identifying information about you will be disclosed, and any quotes used will remain anonymous. I also want to remind you that your participation is entirely voluntary. You have the right to decline to answer any question, and you can end the interview at any point if you choose.

%\noindent Before we begin, you indicated that you would be able to provide us proof of your experience. Can I briefly see (via screen-share) some examples of chats, or other proof? Note that I will not record, screenshot or otherwise collect information about this. This is simply to verify your eligibility.

\subsection*{Demographics}
\begin{enumerate}
    \item How old are you?
    \item What gender do you identify as? 
    \item Where are you currently located (state and country)?
    \item What is your highest level of education?
    \item What domain or field was your degree (as described above) in?
    \item What is your current employment status?
    \item What is your current marital status?
\end{enumerate}

\subsection*{Scam Anatomy}
\begin{enumerate}
    \item Can you describe in your own words, your first encounter with what you now recognize as a pig-butchering scam?
    \item How did the scammer initiate contact with you? 
    \item What was the nature of your relationship with the scammer? 
    \item What platform(s) did the scammer use to communicate with you? 
    \item What were some things that the scammer said or did, that convinced you to invest in their ventures?
    \item How long did your interactions with the scammer last? 
\end{enumerate}

\subsection*{Impact Assessment}
\begin{enumerate}
    \item How much money did you invest overall in the scammer's venture?
    \item Were you able to recover any of the investment? 
    \item How has the experience you underwent affected your emotional well-being? 
    \item How has the experience you underwent affected your financial situation?
\end{enumerate}

\subsection*{Support Systems}
\begin{enumerate}
    \item Did you report your experience to law enforcement? If not, why? 
    \item If you did report it, what was your experience like, and what was the outcome? 
    \item Did you discuss your experience with friends, family or any other support groups? If not, why?
\end{enumerate}

\subsection*{Prevention}
\begin{enumerate}
    \item Looking back, what were some ``red flags" that you ignored, but just recognize that now? 
    \item What steps are you taking to protect yourself from such scams in the future? 
    \item Is there anything else that I have not asked, but that you feel is important for me to know about your experience? 
\end{enumerate}

\vspace{0.45pt}

\subsection*{\\Conclusion}
\noindent This concludes our interview for today. Thank you for your time and willingness to share your experience. Your Amazon gift card will be sent to you via the email you provided shortly after this call. Please do not hesitate to contact us if you have any questions.

\section{Detailed Lifecycle of Pig-Butchering Scams}
To synthesize our findings, we present Table~\ref{table:lifecycle}, which outlines the seven-stage lifecycle of pig-butchering scams as reported by victims in our study. Each stage captures a distinct phase in the scam’s progression—from initial contact to emotional manipulation, financial exploitation, and post-scam re-engagement. 
\begin{table*}[h!]
\centering
\begin{tabular}{>{\centering\arraybackslash}m{3cm} p{3cm} p{8cm}}
\hline
\textbf{Phase} & \textbf{Duration} & \textbf{Strategies Used} \\
\hline
\textbf{The Lure} & 
Point In Time & 
Unsolicited messages pretending wrong number ($n=18$, 69.2\%) \newline
Approached via dating apps ($n=6$, 23\%) \newline
Approached via social media ($n=2$, 7.7\%) \\
\hline
\textbf{The Bond} & 
3 to 11 months & 
Developed romantic relationships ($n=16$, 61.5\%) \newline
Developed business partnerships ($n=10$, 38.5\%) \newline
Shared personal stories ($n=8$, 30.7\%) \newline
Discussed family and career ($n=11$, 42.3\%) \newline
Showed empathy ($n=5$, 19.2\%) \newline
Mirroring manipulation ($n=9$, 34.6\%) \newline
Sent frequent pictures ($n=17$, 65.4\%) \newline
Audio calls ($n=7$, 26.9\%) \newline
Video calls ($n=3$, 11.5\%) \\
\hline
\textbf{The Bait} & 
1 Day to 2 Weeks & 
Introduced crypto as income source ($n=15$, 57.7\%) \newline
Introduced crypto as hobby ($n=3$, 11.5\%) \newline
Claimed insider knowledge ($n=19$, 73.1\%) \newline
Investment consultant managing portfolio ($n=7$, 26.9\%) \newline
Framed as exclusive opportunity ($n=6$, 23\%) \newline
Provided fake profit screenshots ($n=22$, 84.6\%) \newline
Showed messages confirming returns ($n=5$, 19.2\%) \newline
Provided contacts of supposed investors ($n=4$, 15.4\%) \newline
Emphasized exclusivity and FOMO ($n=18$, 69.2\%) \\
\hline
\textbf{The Feed} & 
2 Weeks to 5 months & 
Access to fraudulent platforms (all participants) \newline
Assisted in account setup ($n=11$, 42.3\%) \newline
Platforms had help/support ($n=5$, 19.2\%) \newline
Allowed small withdrawals ($n=3$, 11.5\%) \newline
Encouraged reinvestment of gains ($n=23$, 88.5\%) \newline
Suggested larger investments ($n=13$, 50\%) \newline
Manipulated platform to appear co-investing ($n=14$, 53.8\%) \newline
Introduced complex strategies ($n=8$, 30.7\%) \newline
Discouraged discussing with others \\
\hline
\textbf{The Squeeze} & 
0 - 2 Weeks; until victim refuses further investment & 
Presented time-sensitive opportunities (all participants) \newline
Claimed insider deals access ($n=12$, 46.2\%) \newline
Claimed ICOs about to explode ($n=9$, 34.6\%) \newline
Increased urgency of communication (all participants) \newline
Emotional manipulation and disappointment ($n=11$, 42.3\%) \newline
Encouraged using credit to invest ($n=9$, 34.6\%) \\
\hline
\textbf{The Cut} & 
Immediate after last investment & 
Block access to accounts ($n=21$, 80.8\%) \newline
Accounts under review/errors ($n=10$, 38.5\%) \newline
Requested fees/taxes for withdrawal ($n=16$, 61.5\%) \newline
Ceased all contact ($n=25$, 96.1\%) \newline
Scammer profiles disappeared ($n=18$, 69.2\%) \\
\hline
\textbf{The Encore} & 
2 Weeks to 3 Months & 
Impersonated law enforcement ($n=15$, 57.7\%) \newline
Requested personal info for investigation ($n=8$, 30.8\%) \newline
Claimed funds found, need fees/taxes ($n=9$, 34.6\%) \newline
%Victims who paid in encore scams ($n=15$, 57.7\%) \newline
Recovery scams by specialists ($n=7$, 26.9\%) \newline
Recovery scams by private investigators ($n=2$, 7.7\%) \newline
Recovery scams by legal professionals ($n=5$, 19.2\%) \newline
Legal threats by impersonated lawyers ($n=4$, 15.4\%) \newline
Legal threats by debt collectors ($n=1$, 3.8\%) \newline
Legal threats by regulatory officials ($n=2$, 7.7\%) \\
\hline
\end{tabular}
\newline
\caption{Summary of Pig-Butchering Scam Lifecycle}
\label{table:lifecycle}
\end{table*}

\end{document}